\documentclass{emulateapj}

\begin{document}

\slugcomment{\apj~accepted}

\title{The Abell 85 BCG: a nucleated, core-less galaxy}

%---------------------------------------------------------------------
%---------------------------------------------------------------------

\author{Juan      P.       Madrid\altaffilmark{1,2}      and      Carlos
  J. Donzelli\altaffilmark{3}}

\altaffiltext{1}{Gemini   Observatory,  Southern   Operations  Center,
  Colina El Pino s/n, La Serena, Chile}
  
\altaffiltext{2}{CSIRO, Astronomy and Space Science, PO BOX 76, Epping, NSW 1710, Australia}

\altaffiltext{2}{Instituto  de Astronom\'ia Te\'orica  y Experimental,
  CONICET-UNC,   Laprida  922,   C\'ordoba,   Argentina;  Observatorio
  Astron\'omico de C\'ordoba, UNC, Laprida 854, C\'ordoba, Argentina}

%----------------------------------------------------------------------
%\altaffiltext{4}{The  University  of  Sydney, 44-70  Rosehill  Street,
%  Redfern, NSW 2016, Australia}
%-----------------------------------------------------------------------
%-----------------------------------------------------------------------
%-----------------------------------------------------------------------
                         
\begin{abstract}

New high-resolution  $r$ band imaging of the  brightest cluster galaxy
(BCG)  in Abell  85 (Holm  15A) was  obtained using  the  Gemini Multi
Object Spectrograph.  These  data were taken with the  aim of deriving
an  accurate surface brightness  profile of  the BCG  of Abell  85, in
particular its central region. The new Gemini data show clear evidence
of  a previously  unreported nuclear  emission  that is  evident as  a
distinct  light  excess  in  the  central kiloparsec  of  the  surface
brightness profile. We  find that the light profile  is never flat nor
does it  present a  downward trend towards  the center of  the galaxy.
That is,  the new Gemini data  show a different  physical reality from
the featureless, ``evacuated core''  recently claimed for the Abell 85
BCG.   After  trying  different  models,  we  find  that  the  surface
brightness profile  of the  BCG of Abell  85 is  best fit by  a double
S\'ersic model.

\end{abstract}


\maketitle
%-----------------------------------------------------------------------
%-----------------------------------------------------------------------
%-----------------------------------------------------------------------

\section{Introduction}

Within  the  current  framework  of hierarchical  structure  formation
(e.g.\  White \&  Rees 1978)  galaxy clusters  are formed  through the
successive mergers of galaxies, galaxy groups, and subclusters. Galaxy
clusters form, thus, the  largest gravitationnally bound structures in
the universe.  Interestingly, X-ray  observations have shown that most
of the baryonic mass of galaxy clusters resides not in galaxies but in
their hot intracluster gas (e.g.\ Jones \& Forman 1984)

Abell 85  is a rich  galaxy cluster located  at a redshift of  $z \sim
0.0555$  with 305  confirmed cluster  members (Durret  et  al.\ 1998).
Abell 85  is a bright X-ray  source that has  been extensively studied
using several  X-ray satellites (e.g.\  Markevitch et al.\  1998; Lima
Neto et al.\ 2001; Sivakoff et al.\ 2008). The X-ray emission of Abell
85  testifies   to  an  intense  past  merging   activity  (Durret  et
al.\ 2005). Moreover,  Abell 85 is not fully  relaxed and is currently
merging  with   at  least   two  satellite  subclusters   (Kempner  et
al.\  2002).  The  complex dynamical  state of  Abell 85  was recently
discussed  in great detail  by Ichinohe  et al.\  (2015).  Due  to its
richness, Abell 85 has also been  the target of several studies on the
morphology-density relation (e.g. Fogarty et al.\ 2014).

Located  in the core  of galaxy  clusters, and  formed through  a rich
merger  history, brightest  cluster galaxies  are, in  turn,  the most
massive and  luminous galaxies  in the universe  (De Lucia  \& Blaizot
2007). Recently, L\'opez-Cruz et al.\  (2014) reported that the BCG of
Abell 85 has  the largest galaxy core ever  discovered.  Note that the
BCG of  Abell 85 has also  been identified as Holm  15A. The unusually
large  core in  the surface  brightness profile  of the  Abell  85 BCG
translates into the presence of a supermassive black hole 
with masses above $M_{\bullet} \sim  10^{11} M_{\odot}$. The  mass of the
black hole is obtained by using scaling relations between galaxy cores
and black hole masses (e.g. Kormendy \& Ho 2013).

The results of L\'opez-Cruz et al.\  (2014) were challenged by Bonfini
et al.\ (2015) who find that the Abell 85 BCG does not have a depleted core. 
In fact, Bonfini et al.\ (2015) find that a S\'ersic profile plus an outer 
exponential component provide a good fit to the data.

Galaxy  cores are  defined as  a  relative light  deficit towards  the
nucleus  of the  galaxy compared  to the  inward extrapolation  of the
surface brightness profile of the outer components of the galaxy.  The
physical theory postulated  to explain the presence of  these cores is
the action of binary supermassive black holes that, through three-body
interactions, slingshot away stars in the galactic center (Begelman et
al.\ 1980).

Due to  their possible link to  black holes and  galaxy formation, the
study  of  cores is  an  active field  with  many  authors looking  at
different  theoretical and  observational  aspects. For instance, through  $N$-body
simulations, Milosavljevi\'c \& Merritt  (2001) modeled the decay of a
black hole binary and how it carves galactic cores.
Observationally, a major development in the study of galaxy cores came
with  the  analysis  of   Advanced  Camera  for  Surveys  data.   This
instrument  on-board HST  provided  both superb  resolution and  large
radial  coverage,   key  factors  in  deriving   an  accurate  surface
brightness profile.  Ferrarese et al.\  (2006) use a uniform sample of
100  galaxies in Virgo  imaged with  the ACS  to derive  their surface
brightness profile.  They find that the surface brightness profiles of
most  galaxies are  well fit  by a  S\'ersic (1968)  profile.  Earlier
studies of  galaxy cores  with the  HST include the  work of  Faber et
al.\ (1997) and Laine et al.\ (2003), among many others.

%---------------------------------------------------------------------
%---------------------------------------------------------------------
\begin{figure}
 \plotone{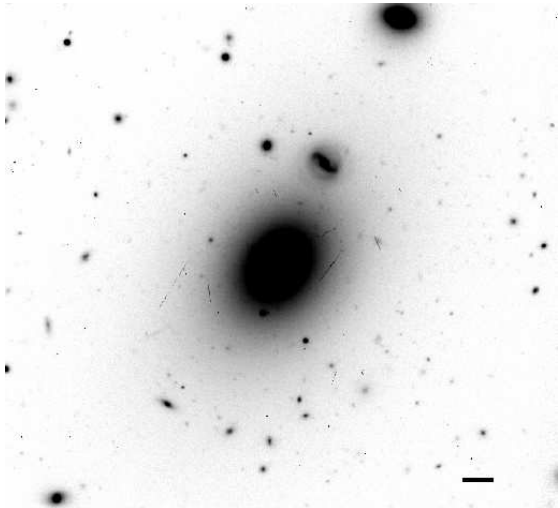}
 \caption{Gemini  Multi Object  Spectrograph image  of Abell  85.  The
   scale bar on the lower right represents a length of 10 kpc. North is
   up and east is left.
 \label{fig1}}
 \end{figure}

%---------------------------------------------------------------------
%---------------------------------------------------------------------
\begin{figure*}
\epsscale{0.7}
 \plotone{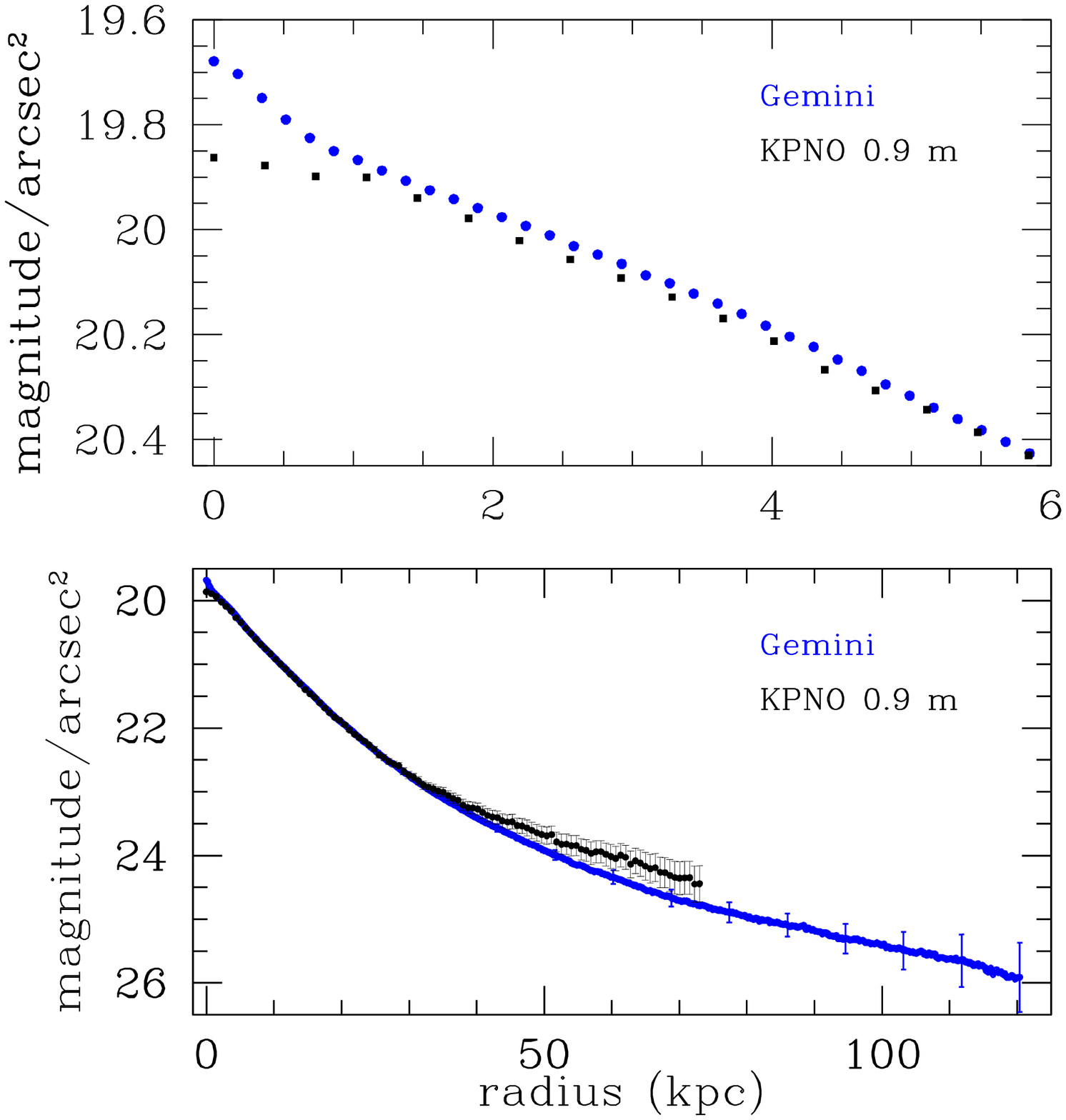}
 \caption{Surface brightness profile of the Abell 85 BCG based on our
   new Gemini data and the KPNO 0.9 m data published by L\'opez-Cruz
   et al.\ (2014).  {\it Top panel:} Zoom of the inner 6 kpc.  The new
   Gemini data shows nuclear emission not present in the KPNO 0.9 m
   data analyzed by L\'opez-Cruz et al.\ (2014).  {\it Bottom panel:}
   Gemini and KPNO 0.9 m data up to 120 kpc in radius.  The surface
   brightness profile of the KPNO data becomes noisy beyond $\sim$40
   kpc. Our new Gemini data goes two magnitudes fainter than the
   KPNO data before reaching similar noise levels.
 \label{fig1}}
 \end{figure*}

%---------------------------------------------------------------------

In   the  following   sections  we   present   high-resolution  Gemini
observations that have  been obtained in order to  study in detail the
surface brightness profile of  the Abell 85 BCG, particularly focusing
on its nuclear region.\\
%---------------------------------------------------------------------

\section{Gemini Observations of Abell 85}

Gemini  South  observations  of  Abell  85  were  obtained  under  the
Director's  Discretionary Time  program GS-2014B-DD-6.   Abell  85 was
observed with  the Gemini Multi Object Spectrograph  (GMOS) on imaging
mode with the detector centered on the BCG, as shown in Figure 1.  The
data  was  obtained  on  2014  November 15  under  stable  atmospheric
conditions on Cerro Pach\'on with a seeing of 0.56 arcseconds.

Two  exposures of  200 s  each were  acquired during  that  night. The
filter in  use was $r\_G0326$, this  filter is centered at  630 nm and
has a filter width of 136 nm. We used a $2\times 2$ binning that gives
an effective pixel scale of 0.160 arcseconds per pixel.

We adopt a redshift of Abell  85 $z\sim 0.0555$ which yields a distance
of 233.9 Mpc  and a scale of 1.075 kpc/arcsec.   The Gemini GMOS pixel
scale for this observation of Abell 85 is thus 172 pc/pixel.

%---------------------------------------------------------------------
%---------------------------------------------------------------------
\section{Data Reduction and Analysis}

Data were processed with the  standard Gemini {\sc PyRAF} package using
the tasks  described in this  section. We obtained bias,  and twilight
flats  from  the observatory.  These  calibration  files were  already
processed through the tasks {\sc gbias} and {\sc giflat}.  The science
images were  bias subtracted  and divided by  the flatfield  using the
task  {\sc gireduce}.   These data  were  acquired using  the new  CCD
detectors  (Hamamatsu) recently  installed on  GMOS. Raw  science data
files have  12 extensions  reflecting the fact  that the  detector has
three CCDs and  four amplifiers per CCD. A  single component image was
made for each  of the two exposures using the  task {\sc gmosaic}. The
final science image was created  combining the two exposures using the
task {\sc imcoadd}.

For the study of the  surface brightness profile, the task {\sc lucy},
within {\sc stsdas},  is used to deconvolve the  image by applying the
Lucy-Richardson algorithm (Lucy 1974; Richardson 1972). The task
{\sc lucy} converges after nine iterations yielding a final resolution 
of  0.45$\arcsec$. 

%---------------------------------------------------------------------

The  {\sc  ellipse} routine  (Jedrzejeswki  1987)  is  applied to  the
science image  in order  to extract the  1D luminosity profile  of the
Abell 85 BCG.  This profile,  directly obtained from the science image
is shown in Figure 2 as blue circles.

In order to obtain an accurate light profile for the targeted galaxy, 
all nearby sources are properly masked and the task  {\sc  ellipse}
ran iteratively until the surface brightness profile converges. We also ran the task {\sc  ellipse} 
with different initial parameters, while we held some (or all) of the parameters 
(center, position angle, and ellipticity) fixed to a constant value to test wether we observe  possible variations of the 
luminosity profile. The resulting luminosity profile turned out to be robust
and shows no dependence on the initial parameters. Also, we did not observe 
any shift in the center of the isophotes during this experiment. Ellipticity remained mainly 
constant i.e. $0.05 < e < 0.1.$ within the innermost few arcseconds.
Small ellipticity values imply relatively large errors for the position angle.

Appropriately  removing  the  sky  background  is also  of  great
importance in order to  obtain an accurate surface brightness profile.
The GMOS  imager provides a relatively  large field of  view, at least
when  compared  to HST  detectors,  this allows  us  to  make a  first
estimate of  the sky background in  an area of the  detector where the
Abell   85 BCG  has  very  low  emission.    

Data taken by the  Canada-France-Hawaii-Telescope (CFHT) and surface 
brightness profile of this galaxy published by Donzelli et al.\ (2011) were also used
to estimate  the sky background.   CFHT data of  Abell 85, in  the $r$
band,  taken under  the  Multi-Epoch Nearby  Cluster Survey  (MENeaCS)
(Sand et al.\ 2011) in September 2008.  The basic assumption we use is
that the  surface brightness  profile of the  BCG of Abell  85
should be identical at intermediate radii, independent of the telescope
in use.

Once a  correct estimate  of the background  is made, both  Gemini and
CFHT surface brightness profiles agree well, with the exception of the
galaxy core where seeing effects dominate.

%---------------------------------------------------------------------
%---------------------------------------------------------------------
%---------------------------------------------------------------------
\begin{figure}
 \plotone{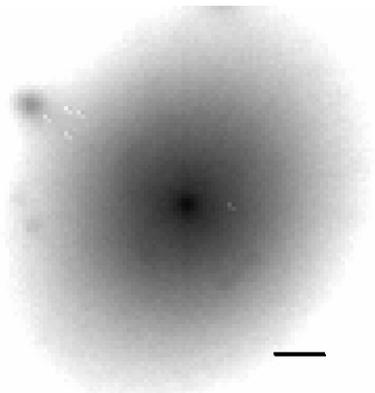}
 \caption{Zoom  of the  Gemini-GMOS image  of the  center of  Abell 85
   showing  a clear  nuclear  emission  that is  seen  on the  surface
   brightness profile as  a clear bump in the  central kiloparsec, see
   top panel of Figure 2. The  scale bar on the lower right represents
   a length of 2 kpc. This image was created using a logarithmic  scale.
 \label{fig1}}
 \end{figure}

%---------------------------------------------------------------------
%--------------------------------------------------------------------- 

\begin{center}
\begin{deluxetable*}{ccccccccccc}
%\tabletypesize{\Large}
\tablecaption{Fits to the Surface Brightness Profile of the BCG of Abell 85 \label{tbl-1}} 
\tablewidth{0pt} 
\tablehead{\multicolumn{10}{c}{Nuker Fits}\\[1mm]
\hline
\colhead{$\mu_b$} & \colhead{$r_b$} & \colhead{$r_b$} & \colhead{$\alpha$} & \colhead{$\beta$} & 
\colhead{$\gamma$} & \colhead{$r_{\gamma}$} & \colhead{Seeing} &  \colhead{$\chi^2$} &\colhead{Telescope}
& \colhead{Reference}
\\
\colhead{(mag/$\arcsec^2$)} & \colhead{($\arcsec$)} & \colhead{(kpc)} & \colhead{} & \colhead{} & 
\colhead{} & \colhead{(kpc)} & \colhead{($\arcsec$)} & \colhead{} & \colhead{}
& \colhead{}
\\
\colhead{(1)} & \colhead{(2)} & \colhead{(3)} & \colhead{(4)} & \colhead{(5)} & 
\colhead{(6)} & \colhead{(7)} & \colhead{(8)} & \colhead{(9)}& \colhead{(10)}& \colhead{(11)}
}

\startdata

21.78   & 17.21   & 18.48  & 1.24  & 3.33 & 0.0  & 4.57 & 1.67 & ... &KPNO  0.9 m & L\'opez-Cruz et al.\\
22.32   & 19.09   & 20.50  & 1.22  & 3.62 & 0.0  & 4.57 & 0.74 & ... &CFHT  3.5 m & L\'opez-Cruz et al.\\
21.05   & 10.70   & 11.56  & 1.90  & 2.29 & 0.14 & 5.02 & 0.56 & 116 &Gemini 8 m & This work\\
%\hline
%\hline\\
%\multicolumn{10}{c}{Double S\'ersic Fits}\\

\cutinhead{Double S\'ersic Fit}

  $\mu_1$ & $r_{e1}$ & $r_{e1}$ & $\frac{1}{n_1}$ &$\mu_2$ & $r_{e2}$ & $r_{e2}$ & $\frac{1}{n_2}$ & $\chi^2$&Telescope & Reference\\
(mag/$\arcsec^2$) & ($\arcsec$) & (kpc) & & (mag/$\arcsec^2$) & ($\arcsec$) & (kpc) & & & \\
(1) & (2) & (3) & (4) & (5) & (6) & (7) & (8) & (9) & (10)& (11) \\[1mm]
\hline\\

 21.71   & 14.59    & 15.68     & 0.933 & 24.87 &  70.79 & 76.10 & 0.839 & 47 &Gemini 8 m & This work\\

\cutinhead{Single S\'ersic Fit}

  $\mu_1$ & $r_{e1}$ & $r_{e1}$ & $\frac{1}{n_1}$ & ... & ... & ... & ... & $\chi^2$ &Telescope & Reference\\
(mag/$\arcsec^2$) & ($\arcsec$) & (kpc) &  & &  & & & \\
(1) & (2) & (3) & (4) & (5) & (6) & (7) & (8) & (9) & (10) & (11) \\[1mm]
\hline\\

23.24   & 342.3  & 368.0  & 0.13   & ... & ...  & ... & ...  & 117 &Gemini 8 m & This work\\

 \enddata

\tablecomments{Nuker  model  fits --  Column  (1): surface  brightness
  $\mu_b$;  Column (2) break  radius $r_b$  in arcseconds;  Column (3)
  break  radius $r_b$  in kpc;  Column  (4) $\alpha$  power radius  at
  $r_b$;  Column (5)  $\beta$; Column  (6) $\gamma$;  Column  (7) cusp
  radius $r_{\gamma}$ in kpc;  Column (8) seeing in arcseconds; Column
  (9)  goodness of fit; (10) telescope in use;  Column (11)  reference.\\ Double  and single
  S\'ersic model fits -- Column  (1 \& 5): central surface brightness;
  Column (2  \& 6):  effective radius in  arcseconds; Column (3  \& 7)
  effective radius in kpc; Column  (4 \& 8) inverse of S\'ersic index $n$; Column
  (9) goodness of fit; (10) telescope; Column (11) reference.\\}

\end{deluxetable*}
\end{center}
%---------------------------------------------------------------------
%---------------------------------------------------------------------
\section{Comparison with recent work}

L\'opez-Cruz et  al.\ (2014)  use the Nuker  model to fit  the surface
brightness profile of  the BCG of Abell 85 derived  with data taken by
the KPNO  0.9 m telescope and  a seeing of  1.67$\arcsec$.  These data
are not publicy available but were given to us by O. L\'opez-Cruz.

As shown in the top panel of  Figure 2 and in Figure 3, the new Gemini
data  reveals the  presence  of nuclear  emission.   This central  and
distinct  feature is  completely  absent from  the  data presented  by
L\'opez-Cruz et al.\ (2014).   In fact, the surface brightness profile
presented by the authors above  is featureless within the inner 20 kpc
of the center of the galaxy.

Using  HST data,  C\^ot\'e et  al.\ (2006)  clearly  demonstrates that
ground  based data  with poor  seeing underestimates  the  presence of
nucleii in nearby elliptical galaxies.

The  Gemini   data  shows  that   from  $\sim$6  kpc   inwards,  the
extrapolation  of the surface  brightness profile  results in  a light
excess  not a  light deficit  as one  might believe  is the  case when
looking at the  lower quality KPNO data. The  above is true regardless
of  the model  chosen to  fit the  surface brightness  profile  of the
galaxy. It should be noted  that the surface brightness profiles shown
in  Figure 2  were  derived  using data  that  received no  additional
processing beyond basic data reduction. 

In their analysis of the CFHT data Bonfini et al.\ (2015) detect
a ``tiny bump" in the light profile within the inner 0.5$\arcsec$.  Indeed, their core-S\'ersic 
model fits a light excess rather than a light deficit within the inner $0.5\arcsec$.

At the faint end of the profile, shown  in the bottom  panel of Fig.\  2,
the KPNO data  becomes noisy beyond $\sim  40$ kpc  from the center  
of the galaxy.   Similar noise levels are  only present in the  Gemini data 
at $\sim  120$ kpc.

%---------------------------------------------------------------------

\section{Nuclear emission and nuclear variability}

The presence of  a clear light excess within  the innermost kiloparsec
of the BCG of Abell 85 prompts us to discuss its physical origin.  One
might attribute this nuclear emission  to the presence of an AGN given
that BCGs are more likely to host a radio-loud AGN than other galaxies
of similar mass (Best et  al.\ 2007).  The detection of X-ray emission
co-spatial with the galaxy core can be candidly thought to be proof of
the existence  of an AGN.  The  picture for Abell 85  is more complex.
In fact,  Sivakoff, et al.\  (2008) exclude the  BCG from a  census of
active  galactic nuclei  in  Abell  85 given  that  its position  also
corresponds, within a few arcseconds, to the peak X-ray emission 
of the intracluster medium.

AGNs  also have  distinctive  radio  emission and  Abell  85 has  been
observed in the radio (Bagchi et al.\ 1998; Slee et al.\ 2001; Schenck
et al.\ 2014).  Based on the  morphology of the radio emission and its
spectrum the above authors do not find evidence of strong AGN activity
for the BCG of  Abell 85. The radio maps of Abell  85 do not show jets
or  lobes that  are the  clear signatures  of strong  and  current AGN
activity.  On  the contrary, those  radio maps are consistent  with the
presence of radio relics from shocked gas or from a dead radio galaxy,
the  latter   not  obviously  cospacial  with  the   BCG  (Schenck  et
al.\ 2014).

We measured  the flux difference  in the core  of the BCG  between the
CFHT  and Gemini  images.  These  images  were taken  about six  years
apart: 2008 September (CFHT)  and 2014 November (Gemini). Fluxes were
measured within an aperture of 1.5$\arcsec$ for each detector. This is
the  aperture at  which the  integrated magnitudes  of a  point source
converge for  both detectors. We  find $\Delta$mag = 0.10  $\pm$ 0.04,
that is, the nucleus of the BCG has become brighter by 0.10 magnitudes
during  the last  six  years.   This type  of  optical variability  is
suggestive of the  presence of an AGN  in the core of the  BCG.  It is
known  that  all  AGNs  vary  in short  timescales  (e.g.\  Ulrich  et
al.\ 1997).  We  should note that we also  measure the flux difference
in the core of a dozen  random galaxies common to both images and find
no difference above the uncertainty level of 0.02 mag. 

The variability within the  core of the BCG of  Abell 85, discussed above,
hints to the presence of an AGN but can also be of stellar origin particularly 
in a dense nuclear stellar structure. Variability is indeed a defining property of AGNs 
that is often used for their discovery. For instance, Cohen et al.\ (2000) search 
for variable galaxies in the Hubble Ultra Deep Field to investigate the 
presence of AGN and find 45 solid candidates. Those AGN candidates 
show characteristic variability of $\Delta$mag~$\sim$0.01~to~0.8 mag.
On the other hand, dense star clusters are favorite crash sites for
binary stars, cataclysmic variables, and classical novae among other
stellar exotica (Knigge et al.\ 2002). Classical novae have been 
found in extragalactic globular clusters and their erupting 
luminosity is comparable to their entire host (Shara et al.\ 2004).

We  find that the core  of the Abell  85 BCG is resolved  with a FWHM of 
about 0.85$\arcsec$, that  is, about 50\% larger than the FWHM of a  stellar PSF.  
At the distance of Abell 85,  the physical size of the central component
is thus $\sim$0.9~kpc. This  central nuclear component, within  the first
arcsecond, can be easily modeled as a gaussian function with an 
integrated magnitude of $m_r~=~22.37$~mag .

The resolved stellar structure in the core of the Abell 85 BCG has a 
physical size that is too large when compared to nuclear star clusters.
Indeed, nuclear star cluster sizes are of the order of a few parsecs 
(e.g.\ $\sim$3 pc; B\"oker 2010). Also, the largest nuclear structure found 
by C\^ot\'e et al.\ (2006) in their survey of the Virgo Cluster has an effective 
radius of 62 pc. 

Based on its size, a Nuclear Stellar Disk (NSD) is a more compatible 
candidate for the origin of the nuclear emission of the Abell 85 BCG.
For instance, Ledo et al.\ (2010) compile a census of nuclear stellar disks 
in early type galaxies. Several of these NSDs have sizes of a few hundred 
parsecs with two of them having sizes larger than 1 kpc. If the central structure 
of Abell 85 is indeed a Nuclear Stellar Disk it would be among the largest reported so 
far. To give more context we remark that the catalogue of Ledo et al.\ (2010) 
is limited to galaxies within 108 Mpc while Abell 85 is at more than twice this distance. 
Also, the Abell 85 BCG is  brigther than the sample studied by  Ledo et al.\ (2010).

Laine et al.\ (2003) study a sample of BCGs with luminosities and distances 
similar to those of Abell 85 and find the presence of two nuclear stellar disks
(Abell 261 and Abell 1142).  Laine et al.\ (2003) also found an additional seven
BCGs with point-like nucleii that exhibit similar morphology to the Abell 85 BCG.

With the current data, we favor a Nuclear Stellar Disk as the physical 
explanation for the nuclear structure present in the core of the Abell 85 BCG.
The nuclear variability we measure is, however, more likely associated with an
AGN. Additional data points are needed in order to build a better sampled 
light curve and unambiguously identify the origin of the variability.\\

%---------------------------------------------------------------------
\section{Fits to the Surface Brightness Profile Using the New Gemini Data}

Different models used to fit  the radial surface brightness profile of
galaxies can be found in  the  literature. Commonly used analytical
functions are  the S\'ersic (1968) profile, which  is a generalization
of  the de  Vaucouleurs (1948)  and exponential  profiles,  the Moffat
(1969) profile, and the  Gaussian profile.  Models that use additional
parameters to  account for the  parametrization of galaxy cores  are a
blend  of two power  laws (Ferrarese  et al.\  1994), the  Nuker model
(Lauer  1995), and the  core-S\'ersic profile  (Graham et  al.\ 2003).
The King (1966) model is commonly used to fit the radial light profile
of globular clusters and small galaxies.

In this section, the results of fitting different analytical models to
the new  Gemini data are presented.  The results of our  best fits and
the fits of  Lopez-Cruz et al.\ (2014) are shown in  Table 1. The fits
below are applied to the deconvolved image. The fits are carried 
out to a galactocentric distance of 115 kpc, that is, where 
the standard deviation of the sky ($\sim$0.3 mag)
equals the uncertainty  on the galaxy surface brightness.

%---------------------------------------------------------------------

\subsection{de Vaucouleurs}

Schombert (1987)  showed that a  de Vaucouleurs (1948) model  fails to
properly fit the surface brightness profile  of the Abell 85 BCG. A de
Vaucouleurs fit for  this galaxy overestimates the flux  at the center
while it underestimates the  flux in the outskirts.  Schombert (1987),
also showed that this result was also true for several other BCGs.\\
%---------------------------------------------------------------------
%---------------------------------------------------------------------
\begin{figure}
 \plotone{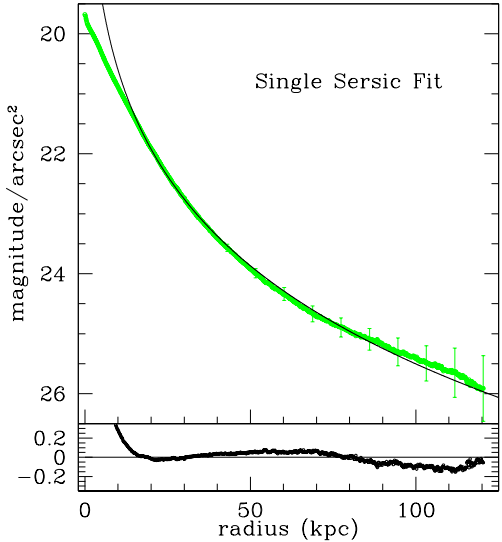}
 \caption{Single  S\'ersic fit  to  the new  Gemini  data (green).   A
   single S\'ersic profile  (solid black line) provides a  good fit to
   the main body of the  surface brightness profile but fails to model
   the data within the inner $\sim$18 kpc. Residuals are shown in the bottom 
   panel.
 \label{fig1}}
 \end{figure}

%---------------------------------------------------------------------
\subsection{Single S\'ersic}

We fit a single S\'ersic profile to the new Gemini data.  For clarity,
we define the S\'ersic profile  in its canonical form $R^\beta$, where
the  concentration parameter  $\beta  =  1/n$ is  the  inverse of  the
S\'ersic index (S\'ersic 1968):

\begin{equation} \label{Ser}
I(r)=I_{e} exp\Big\{-b_n\Big[\Big(\frac{r}r_{e}\Big)^{\beta} - 1\Big]\Big\}
\end{equation}

In  this equation  $I_{e}$ is  the  intensity at  $r =  r_{e}$ at  the
effective radius.  The values for  $b_n$ can be calculated  using $b_n
\sim 2\/n-0.33$ (Caon et al. 1993).

The best  fit of  the S\'ersic model  to the  Gemini data is  shown in
Figure 4.  We find that a single  S\'ersic provides a good  fit to the
data only  over a limited  section of the surface  brightness profile.
Model  and data  diverge  at small  radii,  that is,  for radii  below
$\sim$18 kpc.

The  numerical parameters  of the  best  fit using  a single  S\'ersic
model, such as an exceedingly  large effective radius, expose the fact
that this model does not provide a good physical representation of the
overall surface brightness profile.

Studying  a   large  sample   of  elliptical  galaxies,   Kormendy  et
al.\ (2009) also find that a  single S\'ersic profile is a good fit to
the main section  of the radial profile, while  the model deviates from
the data  at small  radii.  Similarly, Lasker  et al.\ (2014)  need to
include additional  components beyond  a single S\'ersic  profile when
fitting  the  surface  brightness   profile  of  35  nearby  galaxies.
Additional models correspond to  additional physical components such as
bars, nuclei, inner disks, and envelopes.

%---------------------------------------------------------------------

\subsection{Nuker Model}

We also fit  a Nuker model to the Gemini data  and present the results
in   Figure  5.    Numerical  parameters   are  given   in   Table  1.
Interestingly, we  find an even larger  cusp radius, $r_{\gamma}=5.02$
kpc  than the  one  found by  L\'opez-Cruz  et al.\  (2014).  We  find
however a  break radius of $r_b  =11.6$ kpc, almost half  the value of
L\'opez-Cruz et al.\ (2014).

An important  caveat to fitting  a Nuker model  to this galaxy  is the
fact that  it does  not actually  have a flat  evacuated core  and the
Nuker model does  not identify the presence of  the nuclear component.
The existence of this distinct  nuclear component naturally changes the results given 
by the  Nuker model fit, as shown above. Also, it has been proved
that the  Nuker model  is dependent  on the radial  extent of  the fit
(Graham et al.\ 2003). Moreover, the Nuker model was never intended to
fit the  entire surface brightness  profile but the central  region of
any given galaxy -- see the  recent work of Bonfini et al.\ (2015) and
references therein.

The  Nuker model used  by L\'opez-Cruz  et al.\  (2014) underestimates
their data  beyond $\sim$20 kpc. The  failure of the  Nuker profile at
large  radii  prompts  L\'opez-Cruz  et   al.\  (2014)  to  fit  a  de
Vaucouleurs profile (i.e.  S\'ersic profile with $n=4$).  It should be
noted  that  the  de  Vaucouleurs  profile  used  by  L\'opez-Cruz  et
al.\ (2014) overestimates their data in the outskirts of the galaxy.

%---------------------------------------------------------------------
\begin{figure}
 \plotone{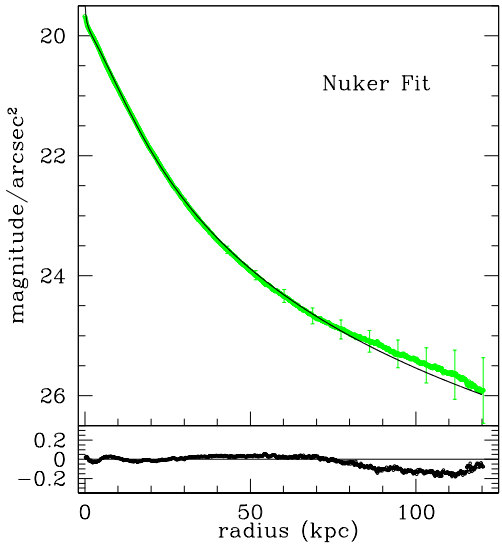}
 \caption{Best Nuker  model fit  (solid line) to  the new  Gemini data
   (green). Residuals are shown in the bottom   panel.
 \label{fig1}}
 \end{figure}

%---------------------------------------------------------------------

\subsection{Core-S\'ersic}

In  a  recent  work  Bonfini  et  al.\ (2015)  carry  out  a  detailed
re-analysis of the  CFHT data of the Abell 85  BCG focusing on fitting
the core-S\'ersic model. Bonfini et al.\ (2015) find that the Abell 85
BCG does not have a depleted core as the light profile does not show a
light deficit when fitted with the core-S\'ersic model. In fact, these
authors  find  that  the Abell  85  BCG  is  not  well adjusted  by  a
core-S\'ersic due to a light  excess in its central surface brightness
profile.   Bonfini et  al.\ (2015)  find  that the  Abell 85  BCG is  a
core-less galaxy  whose surface brightness  profile is best fit  by an
inner S\'ersic profile and an outer exponential halo.

%---------------------------------------------------------------------
\begin{figure}
 \plotone{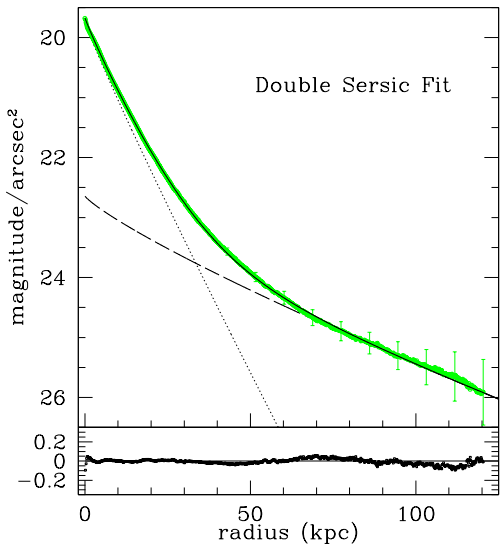}
 \caption{Double S\'ersic  fit to the new Gemini  data (green). Dashed
   and  dotted  lines  represent  the  two inner  and  outer  S\'ersic
   components.  The  solid line,  indistinguishable from the  data, is
   the sum of the two components. Residuals are shown in the bottom 
   panel.
 \label{fig1}}
 \end{figure}

%---------------------------------------------------------------------

%---------------------------------------------------------------------
\begin{figure}
 \plotone{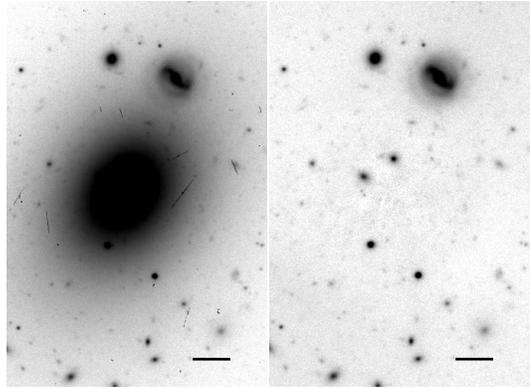}
 \caption{Original image (left) and residual after subtraction of the double S\'ersic model (right).
 No visible structure is left on the residual image. The scale bar on the lower right represents a length of 10 kpc.
 This image was made using a logarithmic display.
 \label{fig1}}
 \end{figure}

%---------------------------------------------------------------------

\subsection{Double S\'ersic}

Gonzalez et al. (2003, 2005) and  Seigar et al. (2007) showed that two
component  models   are  necessary  to  accurately   fit  the  surface
brightness profiles  of BCGs. The  two components pertain to  an inner
and outer component. The existence  of an outer component reflects the
fact  that  BCGs  often  have  extended  envelopes  (Schombert  1987).
Donzelli et  al.\ (2011) studied  the luminosity profiles of  430 BCGs
and found that  about half  of them required  a double S\'ersic  model (S\'ersic $+$ exponential). In
fact,  Donzelli et  al.\ (2011)  fit the  Abell 85  BCG with  an inner
S\'ersic model and an outer exponential component.

We  fit a  double  S\'ersic profile  to  the Gemini  data, and  obtain
satisfactory results. The values of our  best fit are shown in Table 1
and  Figure 6.   One component  has  a S\'ersic  index of  $1/n =  0.93$
(dashed line)  while the second S\'ersic  has an index of  $1/n = 0.84$
(dotted  line).  The sum  of these  two components  (a solid  line) is
indistinguishable from the data.

Models  that explain  the underlying  physical mechanisms  that create
double  S\'ersic   components  have  been  postulated   by  Cooper  et
al.\  (2015).   These authors  postulate  that  for  BCGs, the  double
S\'ersic  profile  originates from  the  superposition  of two  debris
components of  different progenitors. The inner  profile is associated
with relaxed  accreted components while the  outer profile corresponds
with unrelaxed accreted debris (Cooper et al.\ 2015).

The best  fit to the surface brightness  of this galaxy is  given by a
double S\'ersic model.  Values for the goodness of fit  ($\chi^2$) are
presented in Table 1 and a residual image is shown on Figure 7. We should note that the $\chi^2$ value given for the
single S\'ersic fit pertains to a fit between 20 and 115 kpc, a fit over the whole 
range of the data would yield $\chi^2=2750$.

%-----------------------------------------------------------------------------------------------------
%-----------------------------------------------------------------------------------------------------

\section{Does the Abell 85 BCG host the most massive black hole in the Universe?}

Based on the  analysis of new Gemini data  presented above we conclude
that the Abell 85 BCG (Holm 15A) is a nucleated, coreless galaxy. That
is,  it does  not have  an  exceptionally large  core due  to a  light
deficit  in  its central  region  (L\'opez-Cruz  et  al.\ 2014).   Our
results thus nullify  the existence of a supermassive  black hole based
solely on the  presence of a depleted core that  this galaxy, in fact,
does  not have.  

By fitting the Nuker model to the surface brightness profile we find a
large cusp  and break radius.   We refrain from interpreting  the cusp
and break radius as representative of an evacuated core created by the
scouring action of  a binary black hole. A  large cusp radius, derived
from the Nuker model, does not necesarily imply a downward bend of the
inner light profile.

Recently,  Bonfini et al.\  (2015) point  out that  the presence  of a
singular point in  the surface brightness profile of  any given galaxy
does  not imply  the presence  of a  depleted core.   In the  words of
Bonfini et  al.\ (2015) most  galaxies have particular values  for the
negative logarithmic slope of the  intensity profile but this is not a
sufficient condition for the existence of a depleted core.

Moreover, the central brightness profile of the Abell 85 BCG is indeed
different from the flat, or even decreasing surface brightness profile
of, for  instance, the BCG of  Abell 2261 (Postman et  al.\ 2012). The
presence of nuclear structure is difficult to reconcile with a core of
$\sim5$ kpc where other stars within that core are ejected.

%---------------------------------------------------------------------
%---------------------------------------------------------------------
\acknowledgments

The authors would like to thank the referee for a detailed and constructive
report that helped us  improve this paper.
We  are  grateful to  Nancy  Levenson,  Gemini  Head of  Science,  for
granting   us  Director   Discretionary  Time   to  carry   out  these
observations  under program GS-2014B-DD-6.   We also  thank  Gemini observer  Mischa
Schirmer,  and  Gemini  visiting   astronomer  Ricardo  de  Marco  for
obtaining the  data.  The CFHT  data was kindly  shared with us  by Melissa
Graham and David  Sand, likewise, the  KPNO data was made available to us
by Omar Lopez-Cruz. This research  has made use of the NASA Astrophysics
Data  System Bibliographic services  (ADS) and  Google. This  work was
supported by a grant from SECYT-UNC, Argentina.

Based  on observations  obtained at  the Gemini  Observatory  which is
operated by the Association of Universities for Research in Astronomy,
Inc.,  under a cooperative  agreement with  the NSF  on behalf  of the
Gemini partnership:  the National Science  Foundation (United States),
the   National  Research  Council   (Canada),  CONICYT   (Chile),  the
Australian    Research   Council   (Australia),    Minist\'{e}rio   da
Ci\^{e}ncia, Tecnologia e  Inova\c{c}\~{a}o (Brazil) and Ministerio de
Ciencia, Tecnolog\'{i}a e Innovaci\'{o}n Productiva (Argentina). 

%---------------------------------------------------------------------
%---------------------------------------------------------------------

%---------------------------------------------------------------------
%---------------------------------------------------------------------
%---------------------------------------------------------------------

\end{document}